\documentclass[trackchanges,twocolumn]{aastex631}
\usepackage{natbib}
\usepackage{amssymb,amsmath}
\usepackage{graphicx}
\usepackage{multirow}
\usepackage{color}

\newcommand{\uofa}{\affiliation{Lunar and Planetary Laboratory, University of Arizona, Tucson, AZ 85721, USA}}
\newcommand{\princeton}{\affiliation{Department of Astrophysical Sciences, Princeton University, Princeton, NJ 08544, USA}}
\newcommand{\uchicago}{\affiliation{Department of Astronomy and Astrophysics, University of Chicago, Chicago, IL, USA}}
\newcommand{\pppl}{\affiliation{Princeton Plasma Physics Laboratory, Princeton, NJ 08540, USA}}

\usepackage{fancyhdr}

\newcommand{\V}[1]{\mathbf{#1}} 

\usepackage[T1]{fontenc}

\usepackage{float}

\UseRawInputEncoding

\newcommand{\revise}[1]{{\color{black}#1}}

\begin{document}

\title{Electron Influence on the Parallel Proton Firehose Instability in 10-Moment, Multi-Fluid Simulations}

\author[0000-0001-9717-8718]{Jada Walters}\uofa
\author[0000-0001-6038-1923]{Kristopher G. Klein}\uofa
\author[0000-0003-1945-8460]{Emily Lichko}\uchicago
\author[0000-0001-6835-273X]{James Juno}\pppl
\author[0000-0003-0143-951X]{Jason M. TenBarge}\princeton

\begin{abstract}
Instabilities driven by pressure anisotropy play a critical role in modulating the energy transfer in space and astrophysical plasmas. 
For the first time, we simulate the evolution and saturation of the parallel \revise{proton} firehose instability using a multi-fluid model without adding artificial viscosity. 
These simulations are performed using a 10-moment, multi-fluid model with local and gradient relaxation \revise{heat-flux} closures in high-$\beta$ proton-electron plasmas. 
When these higher-order moments are included \revise{and pressure anisotropy is permitted to develop in all species,} we find that the electrons have a significant impact on the saturation of the \revise{parallel} proton firehose instability, modulating the proton \revise{pressure} anisotropy as the instability saturates. 
\revise{Even for lower $\beta$s more relevant to heliospheric plasmas, we observe a pronounced electron energization in simulations using the gradient relaxation closure.}
Our results indicate that resolving the electron pressure anisotropy is important to correctly describe the behavior of multi-species plasma systems.\\
\end{abstract}

\section{Introduction}
\label{sec:intro}

Astrophysical systems, including the local solar environment, contain low-density, magnetized plasmas. 
In these weakly collisional plasmas, non-equilibrium features can develop, in particular, pressure anisotropy. 
Velocity distributions of protons, electrons, and minor ions in the solar wind have all been observed to develop pressure anisotropy \citep{Marsch:2006, Stverak:2008, Hellinger:2006, Verscharen:2013, Huang:2020}. 
Given that high-$\beta$ ($\beta_{j} = 8\pi n_j k T_{j} / B^2$ where $n_j$ and $T_j$ are the density and temperature of a plasma component $j$, $k$ is the Boltzmann constant, and $B$ is the magnetic field) plasma environments such as black-hole accretion flows and galaxy clusters are known to contain winds and regions of expanding and compressing flows, pressure anisotropy instabilities likely play a role in the dynamics of astrophysical systems \citep{Sharma:2006, Schekochihin:2006, Rosin:2011, Riquelme:2012, Kunz:2018}.

Kinetic instabilities (e.g. the Alfv\'en-ion cyclotron, mirror, and firehose instabilities) constrain the development of pressure anisotropy in these systems, bringing them towards isotropy \citep{Gary:1993}. 
These instabilities are important mechanisms for energy transfer between the electromagnetic fields and the particles in weakly collisional plasma systems. 
Firehose instabilities, operating when $P_{\perp} / P_{\parallel} < 1$ (where perpendicular and parallel are defined with respect to the direction of the local magnetic field), are of particular interest as this type of pressure anisotropy is naturally driven in expanding plasma systems \citep{Bale:2009, Matteini:2012}. 
\revise{P}ressure anisotropy-driven instabilities are thought to affect energy transport and evolution of \revise{astrophysical plasma} systems from small to large scales \citep{Kunz:2014, Zhuravleva:2019, Arzamasskiy:2023, Squire:2023}. 

Numerical simulations of firehose instabilities have been performed to characterize pressure regulation in plasma systems such as the solar wind and higher-$\beta$ plasmas. 
However, such simulations typically rely on hybrid descriptions, where the kinetic features of some of the constituent populations are suppressed for computational expediency. 
These simulations have the benefit of being able to describe the anisotropic pressure of one of the species under consideration, but simplifies the other, typically the electrons, to an isothermal, massless neutralizing fluid. 
The fully kinetic approach allows realistic modeling of all plasma species, but kinetic simulations of firehose instabilities are generally limited in system size and focus on just the evolution of electron-scale instabilities \citep{Innocenti:2019} or employ reduced mass ratios to better capture the ion scales \citep{Riquelme:2015} due to the computational cost involved in such models.
The interplay of protons and electrons has been investigated in recent kinetic simulations of the parallel proton firehose instability, with results indicating that electron anisotropy can play an important role in the dynamics of the proton-scale instability \citep{Micera:2020}.
To resolve both \revise{proton} and electron-scale physics in the parallel proton firehose instability for larger simulation boxes, we use a computationally efficient, multi-fluid approach to run multi-scale simulations of high-$\beta$ plasmas with strongly anisotropic initial proton populations.

In this work, the parallel \revise{proton} firehose instability in a proton-electron plasma is simulated using the 10-moment, multi-fluid framework included in the numerical code \texttt{Gkeyll} \citep{Hakim:2006, Hakim:2008}. This framework allows both the proton and electron pressure anisotropies to be resolved. Simulations are run using both the isotropic relaxation \citep{Wang:2015} and gradient relaxation \citep{Ng:2020} closures for the 10-moment system of equations.

\begin{figure}
    \centering
    \includegraphics[width=\linewidth]{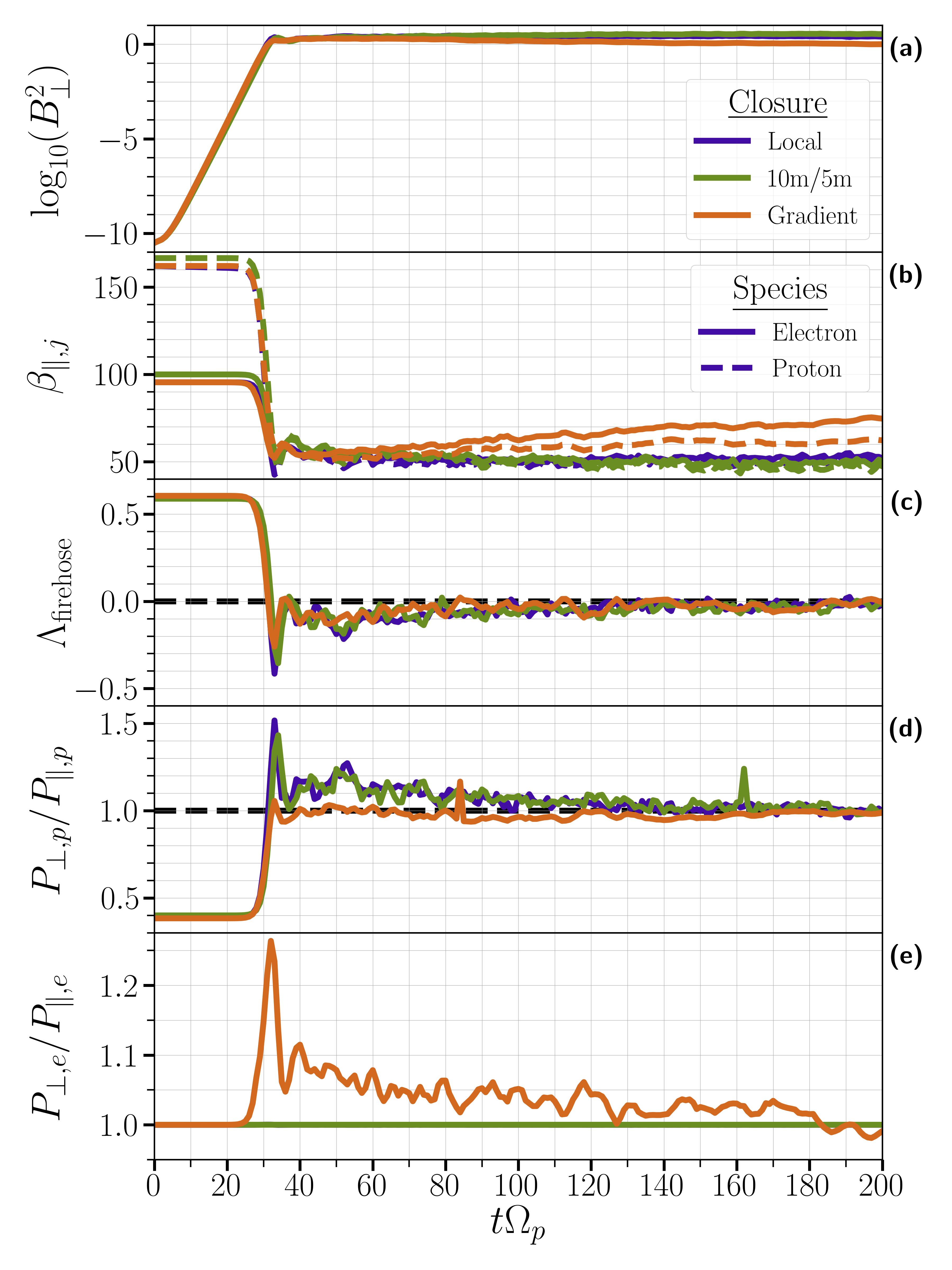}
    \caption{Time evolution of simulations using different heat-flux closures; the gradient relaxation closure from Eqn.~\ref{eq:gradclos} (orange), local relaxation closure from Eqn.~\ref{eq:localclos} (purple), and the 10m/5m model (green). The 10m/5m model is a mixed closure simulation that employs the 10-moment model with the gradient relaxation closure for the protons and the 5-moment model with isotropic scalar pressure for the electrons. Shown in each row are: \textbf{(a)} magnetic field amplitude, \textbf{(b)} parallel plasma $\beta_{\parallel,j}$, \textbf{(c)} firehose instability criterion $\Lambda_{\mathrm{firehose}}$, Eq.~\ref{eq:criterion}, \textbf{(d)} proton pressure anisotropy, and \textbf{(e)} electron pressure anisotropy. Only the gradient relaxation closure simulation develops  electron pressure anisotropy as the parallel proton firehose instability saturates.}
    \label{fig:Closure}
\end{figure}

The paper is organized as follows: In section~\ref{sec:sim}, we describe the 10-moment system of equations and closures and the simulation set-up for our high-$\beta$ proton-electron simulations of the parallel firehose instability. Section~\ref{sec:res} describes the results of our parallel \revise{proton} firehose simulations, including a convergence resolution study\revise{, comparisons to linear kinetic theory, and simulations initialized at different proton pressure anisotropies and proton $\beta$s}.
In Section~\ref{sec:dis}, we summarize our results.

\section{Simulation Setup}
\label{sec:sim}

The 10-moment equations (Eqs.~\ref{eq:1momeq}-\ref{eq:3momeq}) are obtained by taking moments of the Vlasov equation. These equations are solved for each plasma species, so we leave out the species subscript below:
\begin{equation} 
\frac{\partial n}{\partial t} + \frac{\partial}{\partial x_j} \left( n u_j\right) = 0, \label{eq:1momeq} 
\end{equation}
\begin{equation}
m \frac{\partial}{\partial t} \left(n u_i\right) + \frac{\partial \mathcal{P}_{ij}}{\partial x_j} = nq\left( E_j + \epsilon_{ijk} u_j B_k \right), \label{eq:2momeq} 
\end{equation}
\begin{equation}
\frac{\partial \mathcal{P}_{ij}}{\partial t} + \frac{\partial \mathcal{Q}_{ijk}}{\partial x_k} = nqu_{[i}E_{j]} + \frac{q}{m} \epsilon_{[ikl}\mathcal{P}_{kj]} B_l. \label{eq:3momeq}
\end{equation}
In these equations, $q$ and $m$ are the species charge and mass, $\epsilon_{ijk}$ is the three-dimensional Levi-Civita symbol, and $E$ and $B$ are the electric and magnetic fields. The number density $n$, bulk velocity $u_i$, $\mathcal{P}_{ij}$, and $\mathcal{Q}_{ijk}$ are the zeroth, first, second and third order moments of the distribution function. The bracket notation denotes the sum over permutations of the indices, i.e., $u_{[i}E_{j]} = u_iE_j + u_jE_i$ in Eq.~\ref{eq:3momeq}. To solve this system of equations, a closure for the heat-flux tensor, $q_{ijk}$, must be chosen, where the heat-flux tensor and third moment tensor are related by $\mathcal{Q}_{ijk} = q_{ijk} + u_{[i}\mathcal{P}_{jk]} - 2mnu_iu_ju_k$. We solve Eqs. \ref{eq:1momeq} - \ref{eq:3momeq} using the \texttt{Gkeyll} code, which includes the following possible heat-flux closures.

\begin{equation} 
\partial_m q_{ijm} = v_{th} k_{0} (P_{ij} - p\delta_{ij}),
\label{eq:localclos}
\end{equation}
\begin{equation}
q_{ijk} = - \left({v_{th}}/{|k_0|}\right) \chi \partial_{[i}T_{jk]}.
\label{eq:gradclos}
\end{equation} 
Eq.~\ref{eq:localclos} is the \textit{local relaxation} closure as described in \cite{Wang:2015}. In Eq.~\ref{eq:localclos}, $v_{th} = \sqrt{kT/m}$ is the thermal velocity of a given plasma species, $p$ is the scalar, isotropic pressure, and $k_0$ is an effective collisionality prescribed by the user. The local relaxation closure is analogous to collisional relaxation and isotropizes the pressure tensor $P_{ij}$, which is related to second order moment through $\mathcal{P}_{ij} = P_{ij} + nmu_iu_j$. Eq. \ref{eq:gradclos} is the \textit{gradient relaxation} closure \citep{Ng:2020}. $T_{jk}$ is the temperature tensor, $\chi = \sqrt{4/9\pi}$, and in this closure, the user-selected $k_0$ parameter would best be described as an effective conductivity. The gradient relaxation closure is a tensorial generalization of Fick's law for heat transport and thus improves upon the simpler pressure isotropization imposed by the local relaxation closure. We note that when the 10-moment and closure equations are applied to the electron population, the form of the local relaxation closure isotropizes the electrons rapidly due to its dependence on $v_{th,e}$, the electron thermal velocity, whereas the gradient relaxation closure allows electron anisotropy to develop over ion timescales. The gradient relaxation closure thus better models plasma systems where electron dynamics are important, as evidenced by the success of this method in reproducing magnetic reconnection \citep{Ng:2020}.

We conduct two-fluid, proton-electron simulations of the parallel proton firehose instability in 1D using the 10-moment model with both the local relaxation and gradient relaxation closures. We use a real proton-electron mass ratio $m_p / m_e = 1836$ and set the electron Alfv\'en velocity to $v_{A,e} / c = 0.0125$, where $v_{A,j} = B / \sqrt{4 \pi n_j m_j}$ is the Alfv\'en velocity of a plasma species $j$ and $c$ is the speed of light. We choose an initial proton pressure anisotropy $P_{\perp,p} / P_{\parallel,p} = 0.38$, $\beta_p = \beta_e = 300/\pi$, and isotropic electrons ($P_{\perp,e} / P_{\parallel,e} = 1.0$) such that only the parallel proton firehose instability will be strongly driven, starting well above the \revise{total} instability criterion:
\revise{
\begin{equation} 
\Lambda_{\mathrm{firehose}}
= \frac{-2}{\beta_{\parallel, p}} +  \sum_j \left( \frac{\beta_{\parallel,j}}{\beta_{\parallel,p}} \right) \left( 1 - \frac{P_{\perp,j}}{P_{\parallel,j}} \right). 
\label{eq:criterion}
\end{equation} 
}
This criterion, derived from linear fluid theory, indicates an unstable parallel firehose mode when $\Lambda_{\mathrm{firehose}} > 0$ \citep{Yoon:1990}. The simulation system size is $L_x = 299.79~d_p$ (with the x-direction parallel to the mean magnetic field $B_0$) resolved with 4480 cells, where $d_p =  v_{A,p}/\lvert\Omega_p\rvert$ is the proton inertial length and $\Omega_p=q_p B/m_p$ is the proton cyclotron frequency. 
A temporal step of $\Delta t = \Omega_p^{-1}$ is chosen.
For the effective collisionality $k_0$ in the local relaxation closure, we choose $k_{0,p} = 2.74\times10^{-6}$ (a low effective collisionality that allows pressure anisotropy to develop) and $k_{0,e} = 0.1 \sim 1/d_e$. We choose the equivalent values for the effective conductivities $k_{0,\revise{j}}$ in the gradient relaxation closure, which are the inverse of the effective collisionality due to the functional forms of the closures. 
Perturbations of the form $B_{y,z} = -A \sum_{i=0,47} \alpha_i \sin(i k_x x + 2\pi \phi_i)$ are applied to seed the simulation with noise, where $k_x = 2\pi / L_x$, $A = B_0\times10^{-6}$, and $\alpha_i$ and $\phi_i$ are random numbers in $[0,1]$.
We conduct simulations at other spatial resolutions but choose 4480 cells, 0.35 cells per electron inertial length $d_e$, as the fiducial value, capable of capturing electron-scale instabilities.

\begin{figure}
    \centering
    \includegraphics[width=\linewidth]{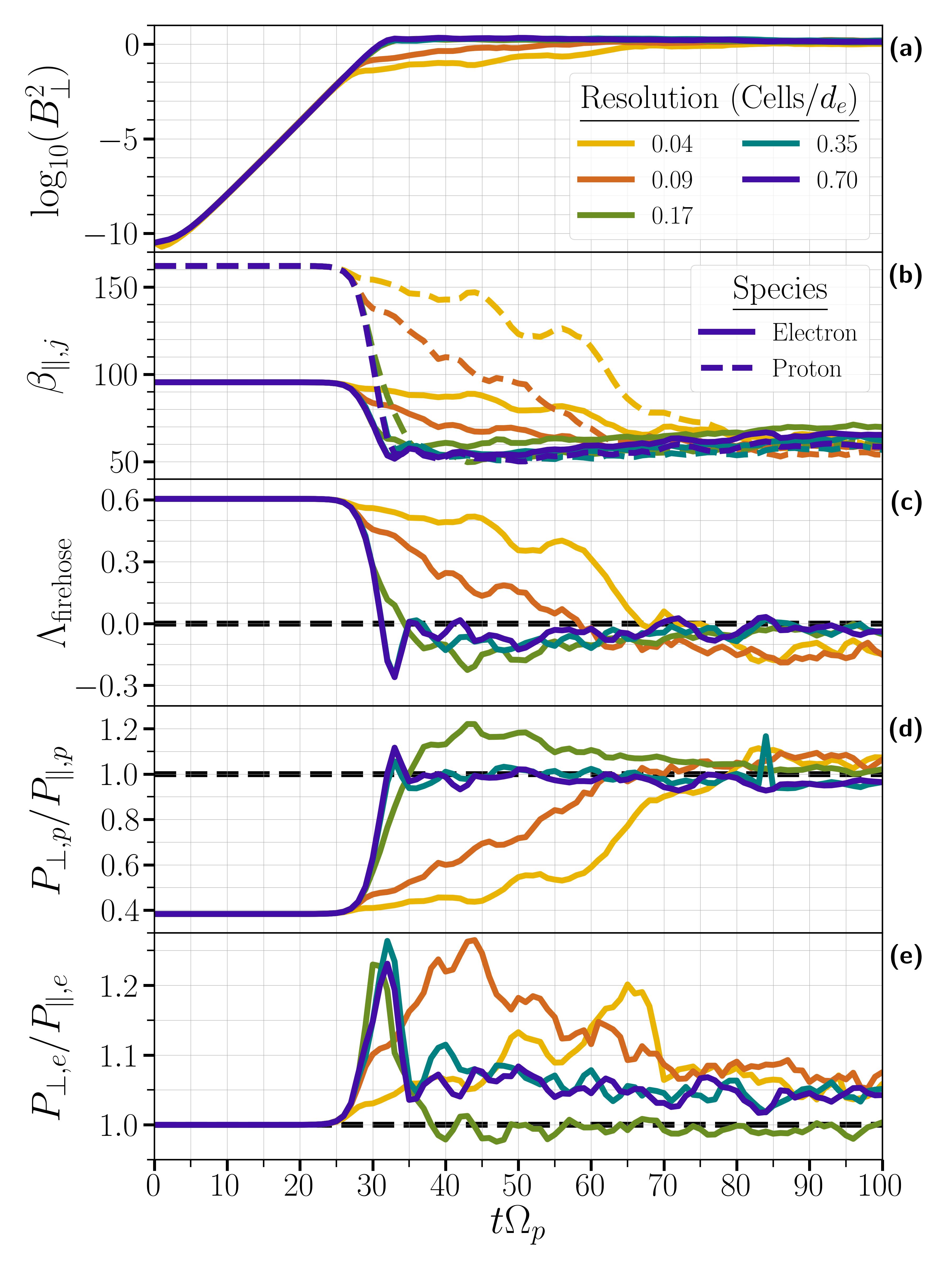}
    \caption{Time evolution of gradient relaxation closure simulations at different spatial resolutions are compared for the same parameters as Fig.~\ref{fig:Closure}. The lowest resolution simulations, 0.04 (yellow) and 0.09 (orange) cells per $d_e$, exhibit secular growth between the linear growth and saturation phases. This secular growth region disappears for the the higher resolution simulations (green, blue, and purple), where the development of electron pressure anisotropy is constrained close to the saturation onset time.}
    \label{fig:Resolution}
\end{figure}

\begin{figure}
    \centering
    \includegraphics[width=\linewidth]{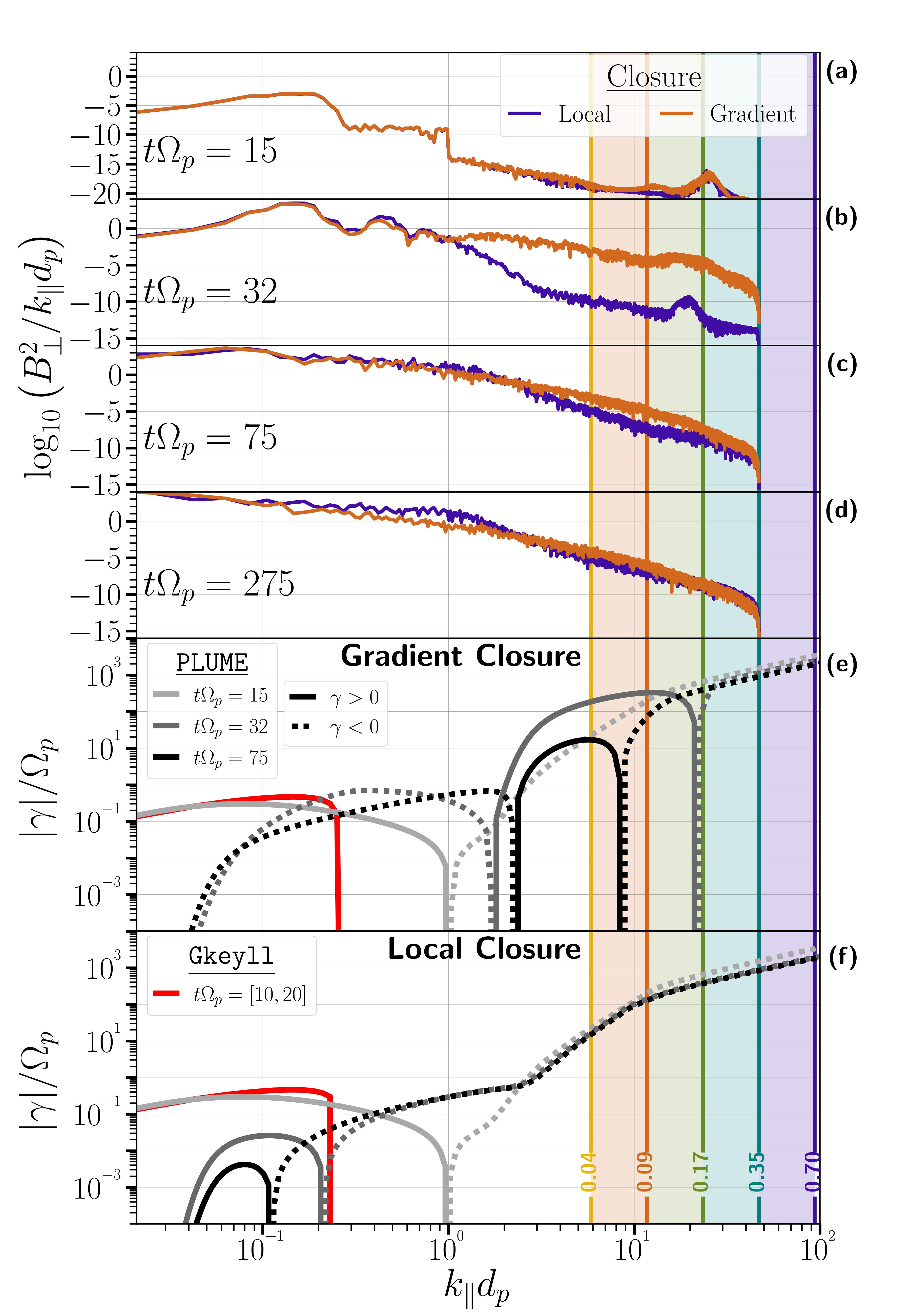}
    \caption{Perpendicular magnetic field data ($B_\perp^2(x,t)$) from the gradient relaxation and local relaxation closure simulations are Fourier transformed to ($B_\perp^2(k_\parallel d_p,t)$) and shown for various simulation times in panels \textbf{(a)} through \textbf{(d)}. Growth/damping rates of instabilities predicted by \texttt{PLUME} using box-averaged plasma parameters at $t\Omega_p = 15$, $32$, and $75$ are shown in gray-scale in panels \textbf{(d)} (for the gradient relaxation closure) and \textbf{(e)} (for the local relaxation closure). Linear instability growth rates extracted from the \texttt{Gkeyll} simulations are plotted in red. The maximum $k_\parallel d_p$ values resolved by the simulations shown in Fig. \ref{fig:Resolution} are indicated by the vertical lines.}
    \label{fig:PLUME}
\end{figure}

\section{Results}
\label{sec:res}

\revise{\subsection{High-$\beta$ Simulations}
\label{ssec:highbeta}
}

We first compare the gradient and local relaxation closure simulations, and we find that both closures are able to drive and saturate the parallel proton firehose instability; a result we are unaware of any other fluid models replicating without introducing unphysical, artificial viscosity such as the hyperresistivity parameter used by \cite{Arnold:2021} in their \texttt{kglobal} fluid-kinetic model.  Fig.~\ref{fig:Closure} (a) shows the linear growth of the parallel firehose instability followed by a saturation at $t\Omega_{p} = 32$. While both closures exhibited similar $B_\perp^2$ growth and saturation, there were distinct  differences in the evolution of the proton and electron pressure anisotropies between the gradient and local relaxation closure simulations. In both simulations, while the initial strong proton anisotropy of $P_{\perp, p} / P_{\parallel, p} = 0.38$ is driven  back to isotropy as the instability saturates (as seen in Fig.~\ref{fig:Closure} (d)), in the local closure case, there is a significant overshoot of the proton anisotropy, which surpasses isotropy and reaches $P_{\perp, p} / P_{\parallel, p} = 1.517$  by $t\Omega_{p} = 33$. As the saturation phase continues, the proton pressure anisotropy stabilizes around isotropy by $t\Omega_{p} = 137$. This behavior \revise{of} the proton pressure anisotropy is not observed in the gradient closure simulation, which barely overshoots isotropy with $P_{\perp, p} / P_{\parallel, p} = 1.0559$ at saturation. Fig.~\ref{fig:Closure} (e) offers an explanation for this difference in the proton pressure anisotropy during the saturation of the parallel firehose instability. While the electrons remain at isotropy throughout the local closure simulation (which is expected due to the speed with which any anisotropy would be damped out by the pressure isotropization imposed by Eq.~\ref{eq:localclos}), the gradient closure allows significant anisotropy to develop in the electrons, which reach $P_{\perp, e} / P_{\parallel, e} = 1.26$ at $t\Omega_{p} = 32$. 

To better understand the influence of the electrons in the gradient relaxation closure simulations, we perform a spatial resolution scan between scales where only the ion physics is resolved to scales where the electron physics is fully resolved. Results are presented in Fig.~\ref{fig:Resolution}. \revise{A resolution of} 0.35 cells per $d_e$ (blue line \revise{in Fig.~\ref{fig:Resolution}}) was used for the simulations in Fig.~\ref{fig:Closure}. While all simulations in the resolution \revise{study} exhibit substantial anisotropy in the electron population as the parallel firehose instability saturates, only for the 0.17 cell per $d_e$ resolution and higher resolution simulations is the anisotropy confined to a narrow time surrounding the saturation at $t\Omega_{p} = 32$. The elevated electron anisotropy observed well past the initial saturation in the 0.04 cells per $d_e$ and 0.09 cells per $d_e$ simulations appears to be linked to under-resolving the modes between the ion and electron inertial lengths. \revise{To properly resolve the electron behavior, we use a resolution of 0.35 cells per $d_e$ for all subsequent results.}

\revise{Using the gradient closure simulation parameters, including the electron pressure anisotropy, we find that linear kinetic theory predicts an electron-scale instability.} In Fig. \ref{fig:PLUME} (e) and (f), we show the growth and damping rates $\gamma$ calculated by \texttt{PLUME}, a numerical Vlasov-Maxwell linear dispersion relation solver \revise{that can model} both species \revise{as} anisotropic bi-Maxwellians \citep{Klein:2015}, for spatially-averaged plasma parameters extracted from the \texttt{Gkeyll} simulations at various times. Initially, an instability is only present in the proton-scale $k_\parallel d_p$ range. For the proton-scale instability, we also plot the growth rate extracted from the \texttt{Gkeyll} simulation data using a logarithmic fit to $B_\perp^2 = A \exp(2\gamma(t - t_0))$ at each value of $k_\parallel d_p$. The \texttt{Gkeyll} growth rate was fit over times $t\Omega_p = 10$ to $20$ to capture the growth associated with the parallel proton firehose instability, and we find that it is in good agreement with the linear kinetic theory results \revise{for both the gradient and local closures. While only this proton-scale instability exists for the local closure parameters, f}or the gradient closure simulation parameters at $t\Omega_{p} = 32$, \texttt{PLUME} finds a second, electron-scale instability that extends to $k_\parallel d_p = 21.10$. The colored vertical lines in Fig.~\ref{fig:PLUME} show the $k_\parallel d_p$ extent resolved by the simulations shown in Fig.~\ref{fig:Resolution}, with the 0.04 and 0.09 cells per $d_e$ simulations cutting off at 5.85 and 11.72 $k_\parallel d_p$ respectively. Thus, only the 0.17 cell per $d_e$ or higher-resolution simulations resolve the $k_\parallel d_p$ range over which an electron-scale instability is predicted by \texttt{PLUME} for the gradient relaxation closure parameters. 
\revise{We identify the \texttt{PLUME} electron-scale instability as the whistler anisotropy instability, a resonant, parallel-propagating fast mode associated with $P_{\perp,e} / P_{\parallel, e} > 1$ \citep{Gary:1993}. 
The frequency and electric field polarization extracted from \texttt{PLUME} are consistent with a forward propagating fast mode.
The cyclotron resonance is the dominant mechanism contributing to the electron growth rate.}

\begin{figure}
    \centering
    \includegraphics[width=\linewidth]{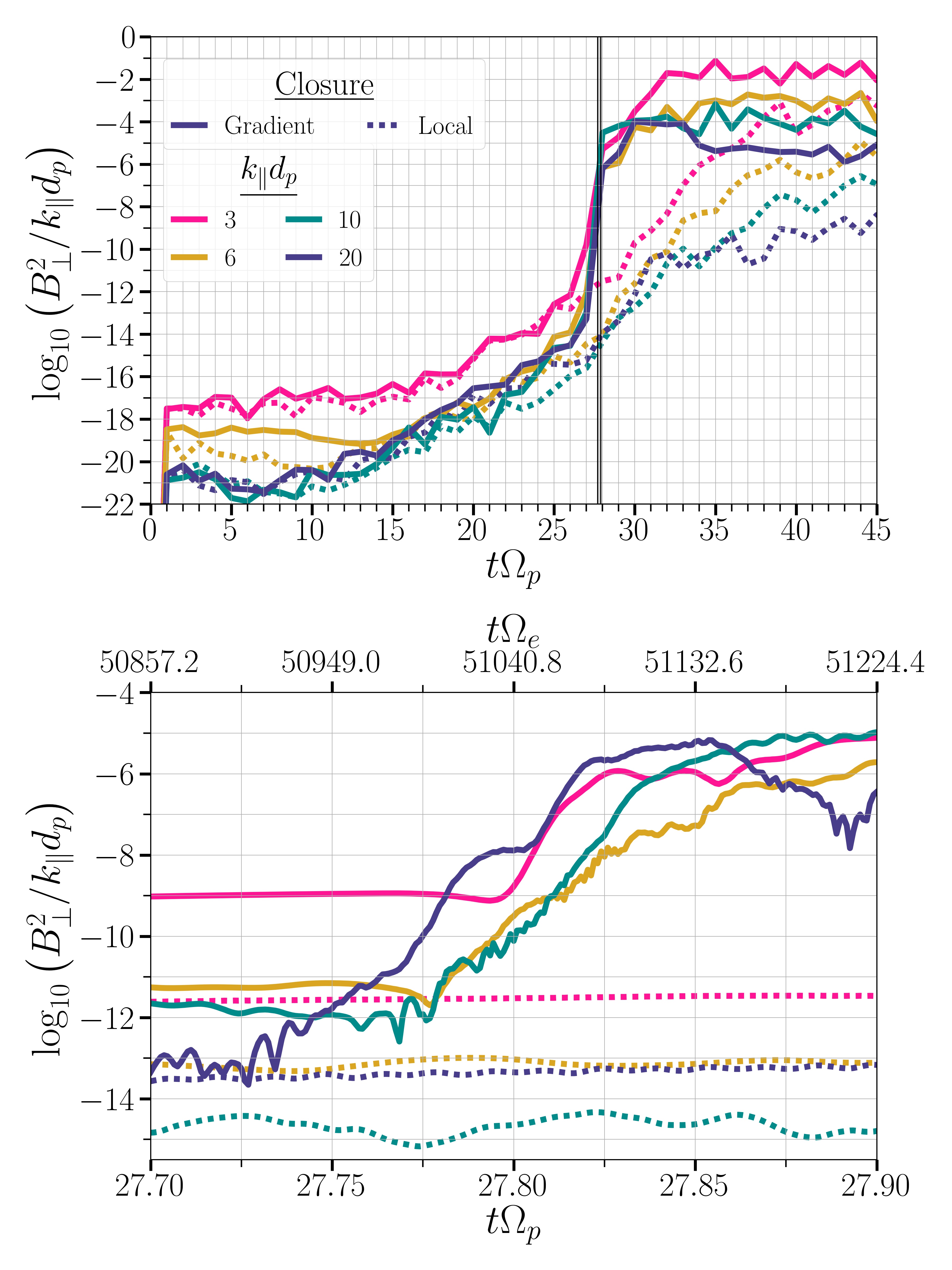}
    \caption{\textbf{Top:} Comparison of perpendicular magnetic field energies at several $k_\parallel d_p$ values for the gradient and local relaxation closure simulations. \textbf{Bottom:} A high cadence output of the same simulations for the  time range $t\Omega_{p} = 27.70$ to $27.90$ (gray highlighted region in top panel). }
    \label{fig:Growth}
\end{figure}

\revise{The absence of the whistler anisotropy instability in the \texttt{PLUME} calculation for the local closure parameters is not the only difference between the two closures that emerges as a result of the strong electron pressure anisotropy in the gradient closure simulation.} The electron dynamics also have a substantial impact on the magnetic field \revise{spectrum of the gradient closure simulation}. The perpendicular magnetic field spectra presented in Fig.~\ref{fig:PLUME} (a) - (d) show enhanced power at small scales in the gradient closure at saturation. As saturation progresses, the spectral peak moves to larger scales for both closures. 
The explosive growth of the magnetic field for $k_\parallel d_p > 1$ in the gradient relaxation closure simulation, which climbs $\approx 8$ orders of magnitude in $0.2~\Omega_{p}^{-1}$, is shown in Fig.~\ref{fig:Growth}. High-cadence ($\Delta t = 1.7~\Omega_e^{-1}$) output from $t\Omega_{p} = 27.70$ to $27.90$ shows little change in the local closure magnetic field for a selection of small-scale $k_\parallel d_p$ over the timescales where electron modes are driven in the gradient closure simulation.

\revise{\subsection{Impact of Electron Pressure Anisotropy}
\label{ssec:elcphys}
}

\revise{ 
To better understand the electron-scale growth observed in the gradient closure simulation, as well as its saturated state, we next examine the physics included in the 10-moment multi-fluid model.
Both the 10-moment model and linear kinetic theory capture the plasma conditions required to drive the whistler anisotropy instability (namely, electron pressure anisotropy), and the linearized 10-moment model contains the first two cyclotron harmonics \citep{Hakim:2008}. 
Calculating the dispersion relation using the linearized 10-moment model yields a proton-scale instability at $t\Omega_{p} = 15$ and electron-scale instabilities at $t\Omega_{p} = 32, 75$ for the same parameters in the Fig.~\ref{fig:PLUME} (e) \texttt{PLUME} calculation. 
However, the 10-moment \texttt{Gkeyll} simulations are fluid, so the mechanism leading to electron-scale magnetic field growth must be something other than resonant wave-particle interactions.

In the gradient closure \texttt{Gkeyll} simulations, we observe a nonlinear response at comparable wavenumbers and growth rates to the \texttt{PLUME} whistler anisotropy instability.
For the gradient closure $B_{\perp}^2$ growth shown in the bottom panel of Fig.~\ref{fig:Growth}, it is possible to construct a rough fit using the same procedure employed for the \texttt{Gkeyll} $\gamma$ fits shown in Fig.~\ref{fig:PLUME} (e) and (f). 
Such a fit yields growth rates on the order of the whistler anisotropy instability calculated by \texttt{PLUME}, $\mid\gamma\mid / \Omega_p \sim 100$, over roughly the same $k_\parallel d_p$ range where the whistler anisotropy instability operates. However, this fit to the electron-scale growth ($k_\parallel d_p > 1.4$) is much worse than the ion instability $\gamma$ fit, with an average $r^2$ statistic of 0.813 (with fits to individual $k_\parallel d_p$ values in many cases having $r^2 < .5$) compared to 0.9997 for the ion-scale growth ($k_\parallel d_p < 0.22$).  
The explosive, but not necessarily linear, $B_{\perp}^2$ growth in \texttt{Gkeyll} at electron scales suggests that the electron-scale modes driven in the gradient closure \texttt{Gkeyll} simulations are associated with nonlinear processes rather than resonant linear processes described by linear kinetic theory.

The question of what processes can transfer energy in fluid simulations also arises when considering the saturation mechanism of the parallel proton firehose instability. While particle scattering has typically been employed as an explanation for pressure isotropization in weakly collisional plasma systems \citep{Sharma:2006, Bale:2009}, our simulations show evidence of magnetic field fluctuations acting as the saturation mechanism. \cite{Rosin:2011} first described magnetic field fluctuations acting as a saturation mechanism for pressure anisotropy-driven instabilities, including the parallel proton firehose instability. In their description, growing fluctuations in the perpendicular magnetic field work against the mean magnetic field to isotropize the pressure. We find evidence of just such a process in our simulations, regardless of closure. Fluctuations in $B_y$ and $B_z$ are initially small, but grow to the order of $B_0$ by the saturation time $t\Omega_p = 32$ and remain large for the remainder of the simulation. 
As saturation progresses, the fluctuations in the gradient closure simulation diminish slightly. A weaker fluctuation strength compared to the local closure is observed by $t\Omega_p = 275$.
The magnetic field spectra in Fig.~\ref{fig:PLUME} (a)-(d) reflect the peak fluctuation energy moving to larger-scale $k_\parallel d_p$ as the saturation progresses. While the growth of perpendicular magnetic field fluctuations contributes to the saturation of the parallel proton firehose instability in both closures, the} electron \revise{pressure} anisotropy and associated explosive growth of the \revise{perpendicular} magnetic field for small-scale $k_\parallel d_p$ values \revise{in only the gradient closure simulations indicate a fundamental difference in how energy is distributed in the plasma for the two heat-flux closures}. 

To confirm that the major difference in the behavior of these closures is whether they allow electron pressure anisotropy to develop, \revise{and not some other difference between the local and gradient closures}, we also \revise{run} a simulation where the 10-moment equations and gradient relaxation heat-flux closure \revise{are} used for the protons and the 5-moment model \revise{is} used for the electrons. The 5-moment model treats the pressure as an isotropic scalar value and is closed by setting the heat-flux vector and viscous stress tensor to 0 as described in \cite{Wang:2015}. This simulation, labeled as the 10m/5m model in Fig.~\ref{fig:Closure}, yields almost identical results to the local closure simulation. This suggests that the proton \revise{behavior is} well-described by both closures, and it is only when we also apply the gradient closure to the electrons that we provide the necessary freedom for the plasma system to deposit energy into the electron population and increase the electron pressure anisotropy.

\begin{figure}
    \centering
    \includegraphics[width=\linewidth]{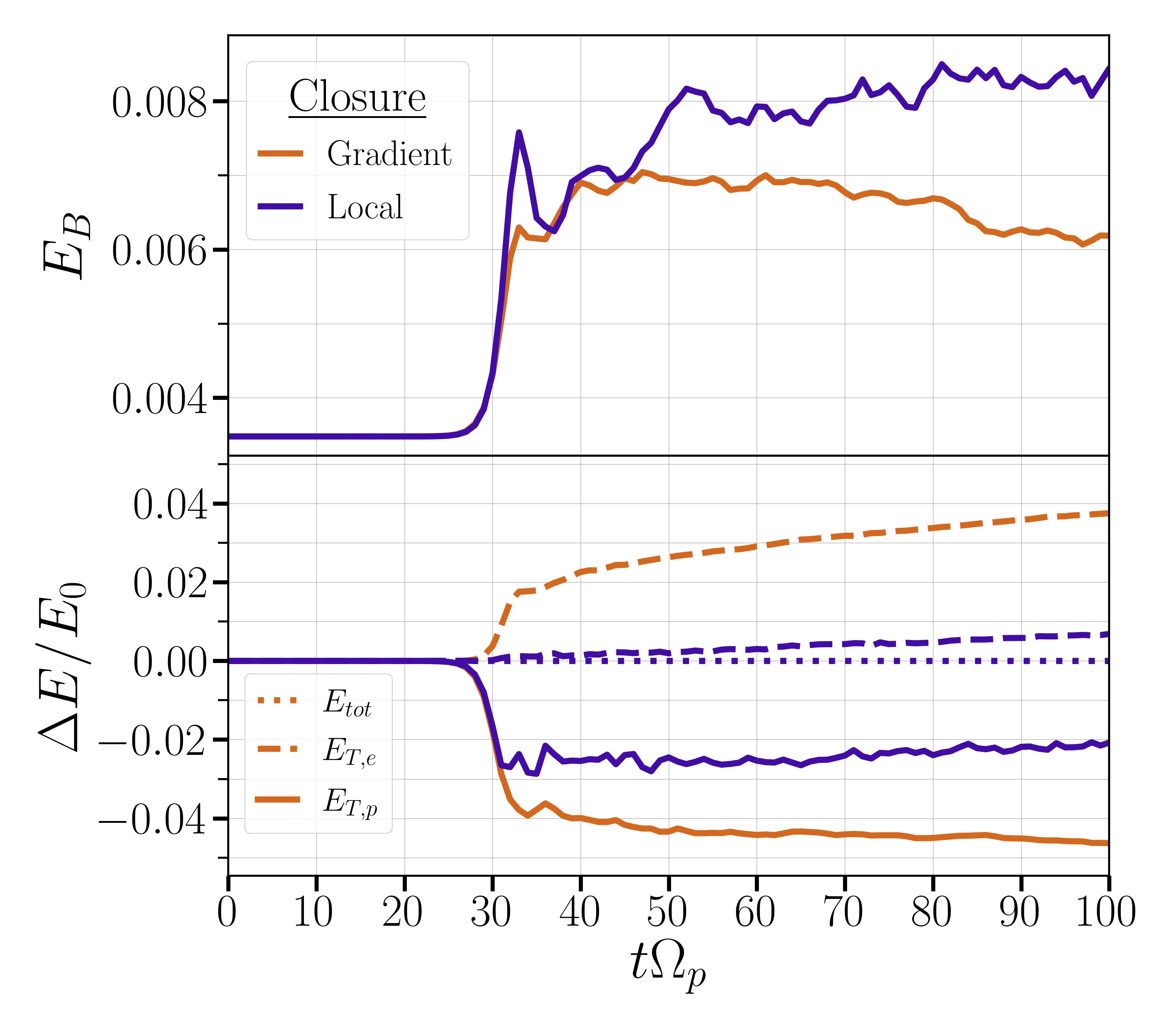}
    \caption{\textbf{Top:} Time evolution of magnetic field energy. \textbf{Bottom:} Change in proton and electron thermal energy and total energy (all normalized to initial energy).}
    \label{fig:Energy}
\end{figure}

The choice of closure results in differing energy partitions as the electrons either remain isotropic or develop strong pressure anisotropy.
Fig.~\ref{fig:Energy} compares the local and gradient closures for magnetic field energy and thermal energy of the electrons and protons (there is negligible drift energy in both cases). At $t\Omega_p = 100$ in the simulation, total particle energy $\sum_j \Delta E_{T,j} / E_0$ in the gradient closure is 3.035 times that of the local closure. The energy is also partitioned very differently between the closures by $t\Omega_p = 100$, with electrons accounting for 45\% of the total thermal particle energy with the gradient closure compared to only 25\% with the local closure. With no electron anisotropy in the local closure simulation, there is little energy transferred to the electrons and more energy is contained in the magnetic field after saturation. \revise{For} the gradient closure \revise{simulation}, both the electrons and the protons represent a significantly larger fraction of the total energy budget.

\revise{The physical picture that emerges from the consideration of the energy partition and the description of the parallel proton firehose instability saturating via magnetic field fluctuations rather than particle scattering is that the electrons present a new energy reservoir that can be used to drive the overall system back to marginal stability.  Ultimately, the plasma saturates at similar $B_\perp^2$ in both closures, but rather than achieving proton pressure isotropization purely through the growth of perpendicular magnetic field fluctuations, the gradient closure allows for energy to be deposited into the electrons to counteract the negative proton pressure anisotropy. We note that this is precluded by the equations developed in \cite{Rosin:2011}, where the electrons are isothermal. Since the magnetic field fluctuations are growing at larger scales than the electron gyroradius, the electrons' magnetic moment, $\mu \sim T_\perp/B$, is well conserved, and thus the electrons adiabatically heat and develop a strong positive pressure anisotropy $P_{\perp, e} / P_{\parallel,e} > 1$. As the instability saturates, the pressure anisotropy of both the protons and electrons contributes to lowering the firehose instability threshold to stability, $\Lambda_{\mathrm{firehose}} \leq 0$. This process accounts for the slight decay in the magnetic field fluctuations seen in the gradient closure compared to the local and 10m/5m closure simulations; as the instability saturates and the proton pressure isotropizes, a portion of the free energy is now allocated to the electron population instead of the magnetic field. 

The development of electron anisotropy also limits the overshoot of proton pressure anisotropy and $\Lambda_{\mathrm{firehose}}$. In Fig.~\ref{fig:Closure} (c) and (d), both the local and 10m/5m closure simulations substantially overshoot stability and isotropy as the proton parallel firehose instability saturates. The electron pressure anisotropy in the gradient closure simulation limits this overshoot, with the proton pressure dropping close to isotropy at the saturation time $t\Omega_p = 32$ and remaining isotropic as saturation progresses. 
Overshoot of proton pressure anisotropy is found in both particle-in-cell and hybrid simulations of shearing plasmas and is generally considered to be a product of finite scale separation between the cyclotron and shearing frequencies in these plasma simulations \citep{Kunz:2014, Riquelme:2012, Squire:2017}. 
That the gradient closure limits this type of overshoot due to the development of finite electron anisotropy may point to the importance of the electron response to more realistically capture the physics of the parallel proton firehose saturation.} This result highlights the potential impact of electron dynamics on the energy transfer and modulation of proton instabilities such as the parallel firehose.

\begin{figure}
    \centering
    \includegraphics[width=\linewidth]{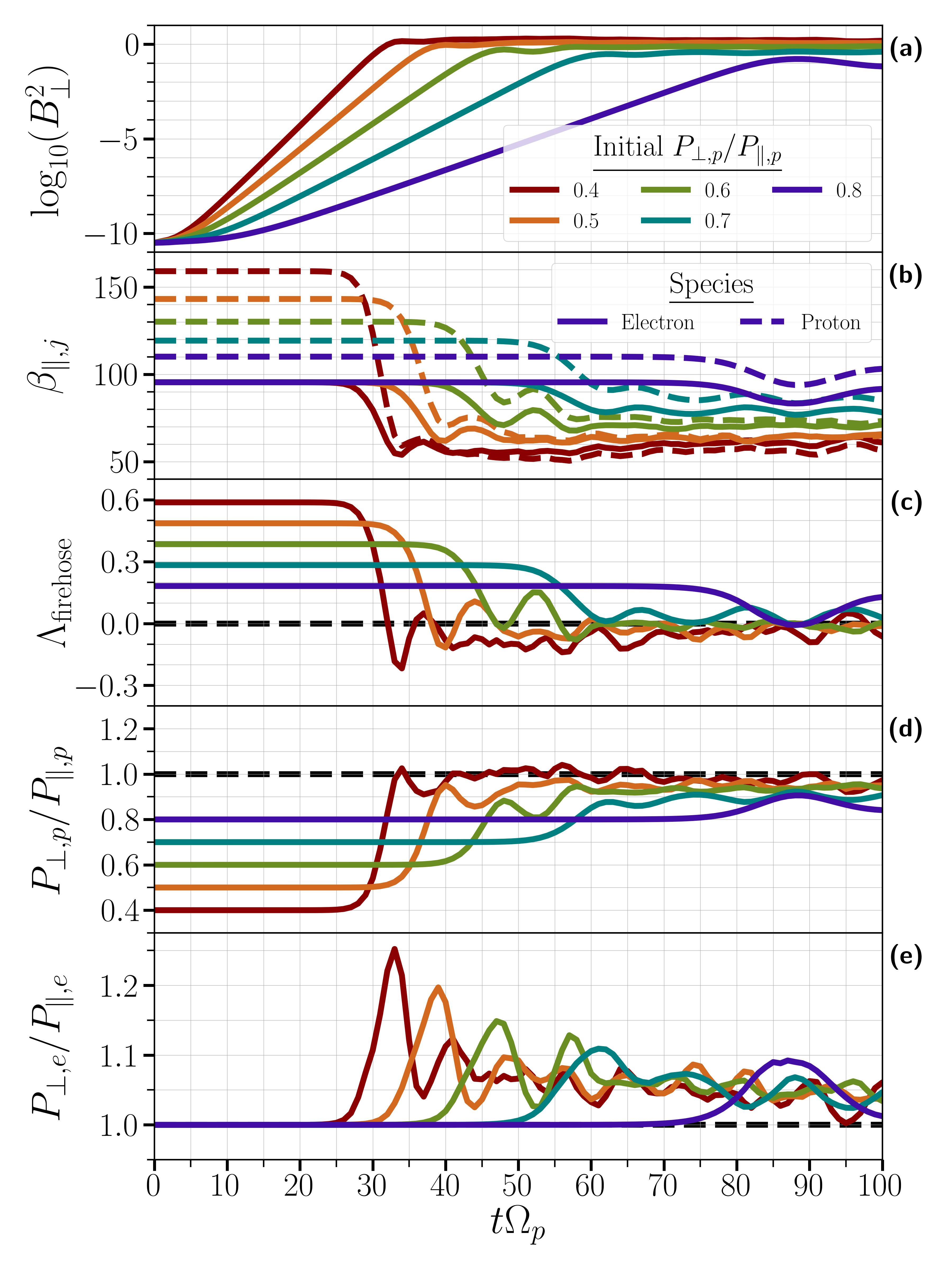}
    \caption{\revise{Gradient closure simulations with different initial proton pressure anisotropies are compared for the same parameters described in Fig.~\ref{fig:Closure}.}}
    \label{fig:VTemp}
\end{figure}

\revise{ \subsection{Varying Initial Pressure Anisotropy}
\label{ssec:pres}

The set of simulations that we have presented above have initial conditions tuned to strongly excite the parallel proton firehose instability with $\beta_p = \beta_e = 300/\pi$ and an aggressive initial proton pressure anisotropy such that $\Lambda_{\mathrm{firehose}} = 0.6$. 
While a $\beta$ in this regime is certainly possible in galaxy clusters and other astrophysical plasmas \citep{Fabian:1994, Peterson:2006, Rosin:2011, Zhuravleva:2019, Kunz:2022}, no observations of proton pressure anisotropy in astrophysical systems are available to inform the choice of initial pressure anisotropy. 
Therefore, we conduct a set of simulations with varying initial proton pressure anisotropies to determine whether the prominent electron response in the gradient relaxation closure simulations presented in Section~\ref{ssec:highbeta} is also present for less strongly excited proton parallel firehose instabilities. For this set of simulations, all of the parameters except the initial proton pressure anisotropy (which is now varied from $P_{\perp,p} / P_{\parallel,p} = 0.4$ to 0.8 in increments of 0.1) are the same as described in Section~\ref{sec:sim}.
A resolution of 0.35 cells per $d_e$ is used in all simulations in order to robustly capture the electron behavior.
Results are presented in Fig.~\ref{fig:VTemp}.

We find that even when the firehose instability criterion $\Lambda_{\mathrm{firehose}}$ is reduced from 0.6 to 0.18 (much closer to stability), electron pressure anisotropy still develops in the system. 
In these less unstable systems, the linear growth phase of the parallel proton firehose instability progresses more gradually, but there is still development of electron pressure anisotropy in all cases as  the linear growth phase ends and the instability begins to saturate.
In Fig.~\ref{fig:VTemp} (e), we observe a decrease in the maximum electron pressure anisotropy as the initial proton pressure anisotropy is reduced.
Even for the least anisotropic case ($P_{\perp,p} / P_{\parallel,p} = 0.8$), the electron pressure anisotropy is still prominent, peaking at $P_{\perp,e} / P_{\parallel,e} = 1.09$.
While the simulation at $P_{\perp,p} / P_{\parallel,p} = 0.4$ and those presented in Section~\ref{ssec:highbeta} have a narrow range of support for the peak electron pressure anisotropy, the time interval over which the development of electron pressure anisotropy is observed broadens to encompass the longer transition from the linear growth phase to saturation in the more weakly unstable parallel proton firehose simulations.
The electron behavior therefore seems to play an important role across a range of proton pressure anisotropies unstable to the parallel proton firehose instability.}

\begin{figure}
    \centering
    \includegraphics[width=\linewidth]{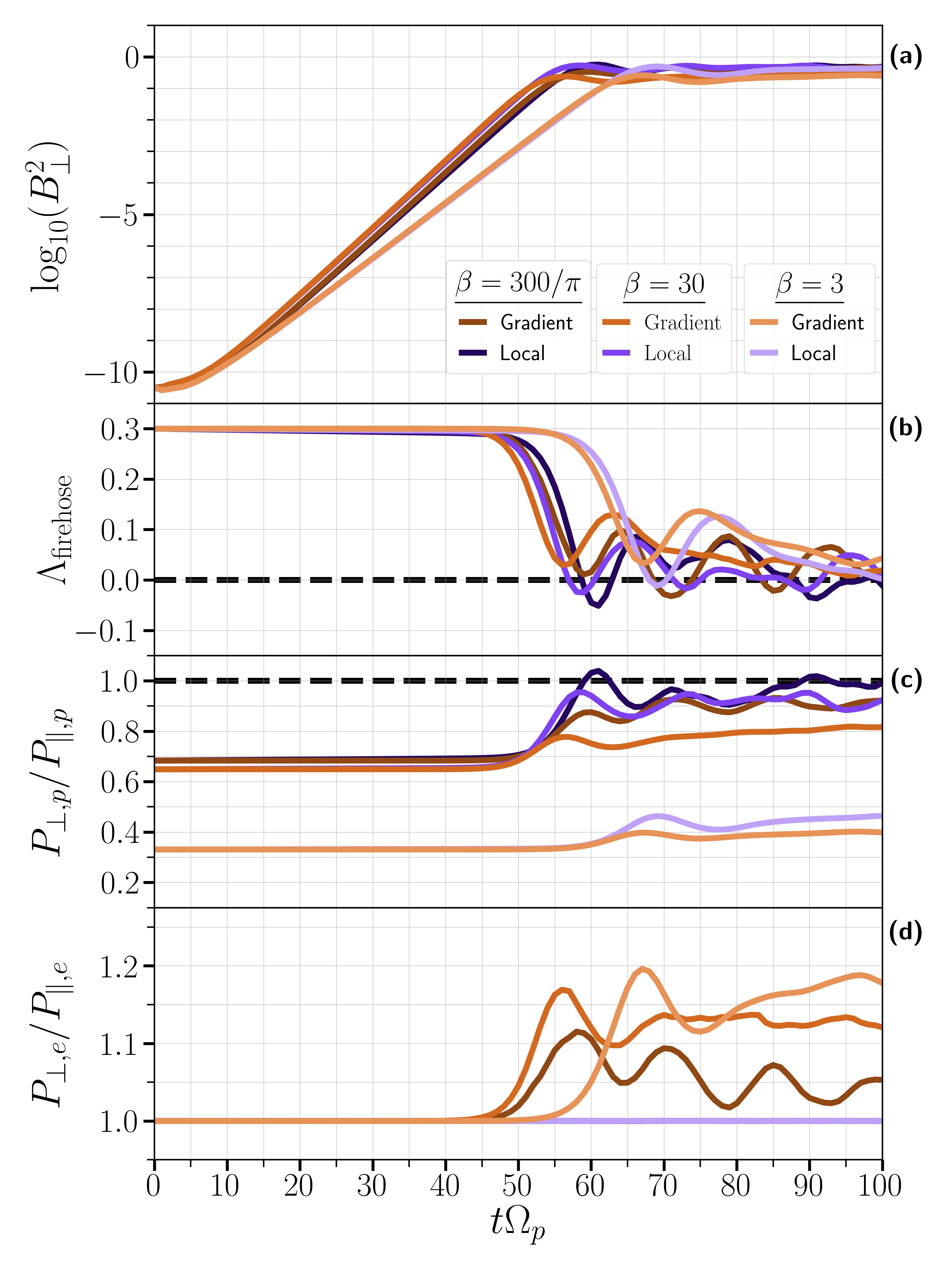}
    \caption{\revise{Gradient and local closure simulations with different initial proton $\beta$s are compared for \textbf{(a)} magnetic field amplitude, \textbf{(b)} firehose instability criterion $\Lambda_{\mathrm{firehose}}$, \textbf{(c)} proton pressure anisotropy, and \textbf{(d)} electron pressure anisotropy.}}
    \label{fig:VBeta}
\end{figure}

\revise{\subsection{A Test of Lower $\beta$s}
\label{ssec:lowbeta}

While we have focused on astrophysical plasma systems with $\beta_p = 300/\pi$ in the above simulations, other weakly collisional space plasmas also tend to develop pressure anisotropies that can trigger pressure anisotropy-driven instabilities such as the parallel proton firehose. 
Plasmas in our solar system, such as the solar wind, generally have lower $\beta$s than their astrophysical counterparts.
At 1 AU, observed solar wind proton distributions peak around $\beta_p \sim 1$ \citep{Wilson:2018}.
As a preliminary exploration of the relevance of our results to lower-$\beta$ plasmas, we present results of simulations at $\beta_p = 300/\pi,$ 30, and 3 initialized with $\Lambda_{\mathrm{firehose}} = 0.3$. 
We choose to fix $\Lambda_{\mathrm{firehose}}$ rather than the pressure anisotropy to simulate similar strength parallel proton firehose instabilities in each $\beta$ regime.

Significant, sustained electron pressure anisotropy develops in the  lower-$\beta$ gradient closure simulations and does not start trending towards isotropy over the simulation run-time of $t\Omega_p = 100$. As in the high-$\beta$ simulations, the local closure develops no electron pressure anisotropy, but the overall magnetic field growth closely matches the equivalent gradient closure simulations.

We also note that we examined but do not show here simulations starting much closer to the instability threshold at $\Lambda_{\mathrm{firehose}} = 0.1$ which still display a weak parallel proton firehose instability.
These weak instabilities have less steep linear growth phases and take longer to saturate.
Our simulations did not achieve a fully saturated state by $t\Omega_p = 200$.
Nonetheless, electron pressure anisotropy began to grow by $t\Omega_p = 130$ for $\beta_p = 300/\pi$ and 30, and by $t\Omega_p = 160$ for $\beta_p = 3$, reaching peak values of 1.05 within the simulation run time.

The simulations presented in Section~\ref{ssec:highbeta} in combination with those shown in Fig.~\ref{fig:VTemp} and Fig.~\ref{fig:VBeta} demonstrate the relevance of this result to space plasmas ranging from astrophysical to heliospheric systems.}

\section{Discussion and Conclusions}
\label{sec:dis}

In this work, we have demonstrated the saturation of the \revise{parallel proton} firehose instability using a 10-moment, multi-fluid (proton and electron) model for two different \revise{heat-flux} closures of the 10-moment system of equations. The closures have differing physical interpretations; the local relaxation closure is analogous to collisional relaxation and the gradient relaxation closure is analogous to Fick's law of diffusion. The local relaxation closure isotropizes the pressure over timescales dependent on the thermal speed of each species. For electrons, this closure works to quickly erase any anisotropy that might develop. The gradient relaxation closure does allow electron anisotropy to develop over proton timescales, and in our simulations, we observe a significant electron anisotropy developing as the parallel proton firehose instability saturates. For parameters extracted from the gradient relaxation simulation at saturation, kinetic linear theory predicts a strong electron-scale instability developing \revise{over} $k_\parallel d_p = 1.81$ to $21.10$. \revise{We identify this instability as the whistler anisotropy instability.} Steep growth in $B_{\perp}^2$ is observed in the simulation at these wavenumbers as the electron pressure anisotropy \revise{grows} at $t\Omega_{p} = 27$, while the local relaxation closure simulation shows no such growth. When allowed to develop, the electron pressure anisotropy plays an important role in the overall energy distribution between the fields and species of a plasma initially tuned to be \revise{strongly unstable} to the parallel proton firehose instability. 
\revise{We find that electron pressure anisotropy develops for not only the extreme initial proton pressure anisotropy of 0.4, but also for simulations of more moderate initial conditions up to $P_{\perp,p} / P_{\parallel,p} = 0.8$.
In addition to the simulations initialized at the astrophysical $\beta_p$ of 300/$\pi$, simulations with lower initial $\beta_p$ ($\beta_p = 3, 30$) that are more relevant to solar wind and other heliospheric contexts also developed electron pressure anisotropy as the parallel proton firehose instability saturated.
These simulations suggest that electron pressure anisotropy may be relevant to many space plasma systems that develop pressure anisotropy, including heliospheric plasmas like the solar wind and magnetospheres as well as the astrophysical plasmas found in accreting black holes and galaxy clusters.

The work presented in this paper is limited to the parallel proton firehose instability and does not consider the influence of oblique modes, including the oblique firehose instability. For the solar wind, the oblique firehose instability may be a stronger constraint on the observed pressure anisotropy \citep{Bale:2009}. Hybrid simulations of expanding plasmas also indicate that oblique firehose modes are dominant over parallel firehose modes in solar wind contexts \citep{Hellinger:2008, Bott:2021}. The dimensionality of our study precludes the development of an oblique firehose mode, and the results presented herein, including the importance of the electron dynamics, might be quite different in a simulation allowing both parallel and oblique firehose modes. 

Kinetic simulations that saturate the parallel proton firehose through pitch-angle particle scattering might also observe different behavior in the development of electron pressure anisotropy. 1D particle-in-cell simulations of the parallel proton firehose instability by \cite{Micera:2020} indicate that electron pressure anisotropy can lead to more quickly-growing parallel proton firehose modes, but in the case where only the proton pressure is initially anisotropic, they find minimal deviation from isotropy in the electron population. Therefore, the saturation mechanism of the firehose instability, either pitch-angle particle scattering or the magnetic field fluctuation mechanism described by \cite{Rosin:2011}, likely plays a critical role in the importance of the electron dynamics in pressure-anisotropic plasmas. 
Pressure anisotropy in astrophysical and space plasma systems is naturally driven by shear or expansion, and a more realistic setup that drives pressure anisotropy through macro-scale dynamics rather than an initialized proton pressure anisotropy is necessary to untangle the potential impact of electron dynamics on space and astrophysical plasmas with the 10-moment model. 
}

With the inclusion of electron pressure anisotropy and realistic closures in the 10-moment, multi-fluid model, our simulations suggest that electrons play a crucial role in mediating the saturation of the proton parallel firehose instability. In particular, our results suggest that the inclusion of more realistic electron physics in simulations of solar and astrophysical plasmas is likely necessary to understand the energy transfer in systems that favor the development of proton instabilities. We emphasize that the 10-moment simulations performed in this study used only modest computational resources, with the simulations at a resolution of 0.35 cells per $d_e$ taking about 3500 CPU hours to run for $100 ~\Omega_{p}^{-1}$, and thus higher-dimensional simulations \revise{that include oblique firehose modes and} a realistic proton-electron mass ratio are well within reach. Future work utilizing this model in higher-dimensional simulations relevant to the solar wind and other astrophysical plasmas where pressure anisotropies are known to exist can provide more insight into the influence of electrons on the ion-scale instabilities that operate in these plasma systems.

\section*{Acknowledgments}
J.W. and K.G.K were supported by NASA grant 80NSSC19K0912. An allocation of computer time through the High Performance Computing (HPC) resources supported by the University of Arizona TRIF, UITS, and Research, Innovation, and Impact (RII) and maintained by the UArizona Research Technologies department is gratefully acknowledged. E.L. was supported by NSF grant 1949802. J. J. was supported by the U.S. Department of Energy under Contract No. DE-AC02-09CH1146 via an LDRD grant. J.M.T. was supported by NASA grant 80NSSC23K0099. The development of Gkeyll was partly funded by the NSF-CSSI program, Award Number 2209471. J. J. and J.M.T. acknowledge the Frontera computing project at the Texas Advanced Computing Center. Frontera is made possible by National Science Foundation (NSF) award OAC-1818253.



\end{document}